\documentclass{aa}
\pdfoutput=1
\usepackage{natbib}                     
\usepackage{amsmath}                    
\usepackage{graphicx}                   
\usepackage[varg]{txfonts}              
\usepackage[english]{babel}             
\usepackage{nameref}                    
\usepackage[normalem]{ulem}             
\usepackage{paralist}                   

\begin{document}

  \title{{Evidence for a spectroscopic direct detection of reflected light from 
    \object{51 Peg b}}\thanks{Based on observations made with ESO Telescopes 
      at the La Silla Paranal Observatory under programme ID  091.C-0271 (with the 
      HARPS spectrograph at the ESO 3.6-m telescope).}\textsuperscript{,}\thanks{The radial velocity data for the HARPS observations of 51 Pegasi are only available in electronic form at the CDS via anonymous ftp to cdsarc.u-strasbg.fr (130.79.128.5) or via {http://cdsweb.u-strasbg.fr/cgi-bin/qcat?J/A+A/} }}

  \author{J. H. C. Martins\inst{1,2,3,6}
          \and N. C. Santos\inst{1,2,3}
          \and P. Figueira\inst{1,2}
          \and J. P. Faria\inst{1,2,3}
          \and M. Montalto\inst{1,2}
          \and I. Boisse\inst{4}
          \and D. Ehrenreich\inst{5}
          \and C. Lovis\inst{5}
          \and M. Mayor\inst{5}
          \and C. Melo\inst{6}
          \and F. Pepe\inst{5}
          \and S. G. Sousa\inst{1,2}
          \and S. Udry\inst{5}
          \and D. Cunha\inst{1,2,3}
         }

  \institute{
          Instituto de Astrof\'isica e Ci\^encias do Espa\c{c}o, Universidade do Porto, CAUP, Rua das Estrelas, 4150-762 Porto, Portugal
          \and
          Centro de Astrof\'{\i}sica, Universidade do Porto, Rua das Estrelas, 4150-762 Porto, Portugal
          \and
          Departamento de F\'isica e Astronomia, Faculdade de Ci\^encias, Universidade do Porto, Rua do Campo Alegre, 4169-007 Porto, Portugal
        \and
        Aix Marseille Universit\'e, CNRS, LAM (Laboratoire d'Astrophysique de Marseille) UMR 7326, 13388, Marseille, France
          \and
          Observatoire de Gen\`eve, Universit\'e de Gen\`eve, 51 ch. des Maillettes, CH-1290 Sauverny, Switzerland
         \and 
        European Southern Observatory, Casilla 19001, Santiago, Chile}

  \date{Received date / Accepted date }
  \abstract{The detection of reflected light from an exoplanet is a difficult technical challenge at optical wavelengths. Even though this signal 
  is expected to replicate the stellar signal, not only is it several orders of magnitude fainter, but it is also hidden among the stellar noise.}
  {We apply a variant of the cross-correlation technique to HARPS observations of 51 Peg  to detect the reflected signal from planet \object{51 Peg b}.} 
  { Our method makes use of the cross-correlation function (CCF) of a binary mask with high-resolution spectra to amplify the minute planetary signal that is present in the spectra by a factor proportional to the number of spectral lines when performing the cross correlation. The resulting cross-correlation functions are then normalized by a stellar template to remove the stellar signal. Carefully selected sections of the resulting normalized CCFs are stacked to increase the planetary signal further. The recovered signal allows probing several of the planetary properties, including its real mass and albedo.}
  {{We detect evidence for the reflected signal from planet \object{51 Peg b} at a significance of 3-$\rm \sigma_{noise}$. The detection of the signal permits us to infer a real mass of $\rm 0.46^{+0.06}_{-0.01}\;M_{Jup}$ (assuming a stellar mass of $\rm 1.04\;M_{\odot}$) for the planet and an orbital inclination of $80^{+10}_{-19}$ degrees. The analysis of the data also allows us to infer a tentative value for the (radius-dependent) geometric albedo of the planet. The results suggest that \object{51Peg b} may be an inflated hot Jupiter with a high albedo (e.g., an albedo of 0.5 yields a radius of $\rm 1.9 \pm 0.3\;R_{Jup}$ for a signal amplitude of $\rm 6.0\pm0.4 \times 10^{-5}$}).}
  {We confirm that the method we perfected can be used to retrieve an exoplanet's reflected signal, even with current observing facilities. The advent of next generation of instruments (e.g. ESPRESSO@VLT-ESO) and observing facilities (e.g. a new generation of ELT telescopes) will yield new opportunities for this type of technique to probe deeper into exoplanets and their atmospheres.}


  \maketitle
  
  -------------------------------------------------
  \section{Introduction}                                        \label{sec:Introduction}
    Since the discovery in 1995 of a planet orbiting 51 Peg \citep{1995Natur.378..355M}, over 1800 planets in around 1100 planetary systems have been found \citep[][]{2011A&A...532A..79S}\footnote{{http://exoplanet.eu/}}: this number increases steadily.  Furthermore, close to 470 multiple planetary systems have been detected, some of which are highly complex \citep[e.g.,][]{2013Sci...340..587B}.
    
    The search for exoplanets has been following two different, but complementary, paths: the detection of exoplanets with increasingly
lower masses, and the characterization of these exoplanets and their atmospheres. On the detection side, the radial velocity {\citep[e.g.,][]{2011exop.book...27L,2013A&A...549A.109B}} and transit methods {\citep[e.g., ][]{2010Sci...327..977B,2011exop.book...55W}} have been the most prolific. One of the most important resul?ts of planet detection surveys is the ubiquity of planets around solar-type stars {\citep[e.g.,][]{2010Sci...330..653H}}.
    
    On the characterization side, the current frontier of exoplanet characterization has been pushed toward the study of exoplanet atmospheres, both from a composition and from a dynamics point of view. To overcome this difficult challenge, several indirect techniques have been developed. Transmission spectroscopy relies on observing the host star spectrum as it is filtered by a planet atmosphere during a transit \citep[e.g.,][]{2002ApJ...568..377C,2014arXiv1403.4602K}. Occultation photometry and spectroscopy measure the occultation (secondary eclipse) depth of the star+planet light curve at different wavelengths to derive the planet's thermal \citep[e.g.,][]{2005Natur.434..740D,2010A&A...513A..76S,2014arXiv1410.2241S} and reflected \citep{2009A&A...501L..23A-eprintv1,2013AN....334..188R} signatures. The detection of the exoplanet emission spectra was also possible through high-resolution spectroscopy \citep{2012Natur.486..502B, 2012ApJ...753L..25R, 2014A&A...561A.150D}. The measurement of phase variations relies on the detection of the flux variation along the planet's orbit as it alternately presents its day and
night hemisphere to us  \citep[e.g.,][]{2009ApJ...703..769K-eprintv1,2011ApJ...740...61K}. These techniques represent the current front line of exoplanet characterization and are limited only by flux measurement precision
that they impose as a result of the low planet-star flux ratio (e.g., in the most favorable cases, for a Jupiter sized planet with a 3 day period orbit $F_{\rm Planet}/F_{Star}\approx10^{-4}$ in the visible and $F_{\rm Planet}/F_{Star}\approx10^{-3}$ in the IR). 
 
    {\cite{1999ApJ...522L.145C} attempted to recover the reflected spectrum of the giant planet orbiting $\tau$ Boo, which paved the way for detecting reflected light in the
optical. To do so, they performed a $\chi^2$ evaluation of simulated star+planet spectra against high-resolution observations obtained
with the HIRES spectrometer at  the Keck observatory. Although unable to detect the reflected signal from planet $\tau$ Boo b, they were able to set an upper limit on the planet's maximum planet-to-star flux ratio of about $5\times10^{-5}$. In the same year, \cite{1999Natur.402..751C} made an attempt at detecting the reflected signal of the same planet using a least-squares deconvolution technique, also without conclusive results. More recently, other attempts with alternative methods have been made \citep[e.g.,][]{2003MNRAS.346L..16L, 2010A&A...514A..23R}, albeit all results have been inconclusive about a definitive detection of a reflected exoplanet signal. Nonetheless, all these attempts are of great scientific values because they allow establishing upper limits on the planet-to-star flux ratio.} 
    
    More recently, researchers were able to collect measurements of the albedo of several exoplanets in an effort to constrain current atmosphere models \citep[e.g.,][]{2011ApJ...729...54C,2014arXiv1405.3798D} and infer their composition \citep[e.g., the presence of clouds in the atmosphere of HD 189733 b as done by ][]{2014arXiv1403.6664B}. An interesting result is the observation of \object{HD 189733 b,} where researchers were able to infer the planet's color from albedo measurements \citep{2013ApJ...772L..16E}. Nonetheless, {several of these} results are still subject to some discussion because it has been shown that the blue excess in the planet's transmission spectrum might also be the result of stellar activity \citep[][]{2014A&A...568A..99O}. 
    
    {The detection of sodium absorption in the transmission spectrum of \object{HD209458} b \citep{2002ApJ...568..377C} paved the way for the detection of spectral features in exoplanet atmospheres. As new facilities were developed and analysis techniques perfected with time, other elements or even molecules were detected  \cite[e.g.,][]{2008ApJ...673L..87R, 2010Natur.465.1049S, 2013MNRAS.436L..35B, 2014ApJ...791L...9M, 2014arXiv1404.3769B, 2014Natur.509...63S}.}
    
    In this paper we apply the technique described in \cite{2013MNRAS.436.1215M} to HARPS observations of 51 Peg {to try to retrieve} the reflected spectrum of its planetary companion. \object{51 Peg} (\object{HD217014}) is a solar-type star, slightly more massive than our Sun \cite[${M_{51\;Peg}}/{M_{\odot}}\approx 1.04$,][]{2013A&A...556A.150S}, at a distance of approximately 15.6 parsec from us \citep{2007A&A...474..653V}. {With a minimum mass slightly under half the mass of Jupiter} and an orbital period slightly longer than days, \object{51 Peg b} is the prototype of a hot Jupiter, giant gas planets in close orbits \citep{1995Natur.378..355M}. The combination of the host brightness \citep[$V_{Mag} = 5.46$,][]{2009ApJ...694.1085V}, the giant planet's large dimensions, and the short-period planetary orbit yield a relatively high planet-star flux ratio and make this planetary system an excellent candidate for testing the method presented in \cite{2013MNRAS.436.1215M}. 
    
    In Sect. \ref{sec:Principle} we describe the principle behind our method. In Sect. \ref{sec:Method} we describe the method and its application to our data. The results are presented in Sect. \ref{sec:Results} and are discussed in Sect. \ref{sec:Discussion}. We conclude in Sect. \ref{sec:Conclusions}.

  \section{Principle behind the method}                               \label{sec:Principle}
    \cite{2013MNRAS.436.1215M} showed that the cross-correlation function (hereafter CCF) can be used to mathematically enhance the S/N of our observations to a level where the extremely low S/N planetary signal can be recovered.  The CCF of a spectrum with a binary mask \citep{1996A&AS..119..373B} has been extensively tested in detecting exoplanets with the radial velocities method. Briefly, this technique corresponds to mapping the degree of similarity between the stellar spectrum and a binary mask (representing the stellar type), which increases the S/N of the data by a factor proportional to the square root of the number of spectral lines identified in the mask.
    
    As discussed in  \cite{2013MNRAS.436.1215M}, we expect the planet's signature to replicate the stellar signal, scaled down as a result of geometric (e.g., planet size) and atmospheric (e.g., albedo) factors. {The planetary albedo measures the fraction of incident light that is reflected by the planet atmosphere. Several albedo definitions exist \cite[e.g.,][]{1999ApJ...513..879M,2002MNRAS.330..187C,2010eapp.book.....S}, but in our study we only considered the geometric albedo $A_{\rm g}$, which is defined as the ratio of a planet's flux measured at opposition ($\alpha = 0$) by the flux of a Lambertian disk at the same distance and with the same radius as the planet. This allows us to define the planet/star flux ratio as
    \begin{equation}
      \frac{F_{planet}(\alpha)}{F_{Star}} = A_{\rm g} \, g(\alpha) \left[\frac {R_{planet}}{a} \right ]^{2}
      \label{eq:FluxesRatio}
    ,\end{equation}
    where $A_{\rm g}$ is the planet's geometric albedo, $\alpha$ the orbital phase, $g(\alpha)$ the phase function, and $R_{planet}$ and $a$ the planetary radius and orbital semi major axis.}.

    
    For \object{51 Peg b}, assuming an albedo of 0.3 and a planetary radius of $1.2\;\rm R_{Jup}$, Eq. \ref{eq:FluxesRatio} will yield a maximum planet-star flux ratio of $\approx\,2\times10^{-5}$.
To detect the planet's reflected signal under these conditions, we consequently need data with a combined S/N of at least $165\,000$ for a 3 $\rm \sigma_{noise}$ detection. A typical G2 stellar CCF-mask used in HARPS has over 4000 spectral lines. This means that in principle a spectrum with a S/N of about 2600 will contain enough information to allow us to build a CCF with the necessary S/N to detect the light spectrum of 51 Peg reflected on its planet (at 3-$\sigma$ level). A lower S/N will suffice if the albedo or the planetary radius are higher (according to Eq. \ref{eq:FluxesRatio}).
    

  \section{Method}\label{sec:Method}
    \subsection{Data}                                       \label{sec:Data}
      Our data were collected with the HARPS spectrograph at ESO's 3.6 m telescope at La Silla-Paranal Observatory as part of the ESO program {091.C-0271}. It consists of {90} spectra observed in seven different nights (\textit{2013-06-08},  \textit{2013-06-25}, \textit{2013-08-02}, \textit{2013-08-04}, \textit{2013-09-05}, \textit{2013-09-09,} and \textit{2013-09-30}), which amounts to about 12.5 h of observing time. These observations were split into several carefully selected time windows in which the planet could be observed close to superior conjunction (i.e., when the day side of the planet faces us) to maximize the planet's flux (maximum phase). These time windows were computed from the ephemeris provided by \cite{2006ApJ...646..505B}.
      
      The obtained  spectra have a S/N on the $\rm 50^{th}$ order ($\sim$5560\AA) that varies between 122 and 388. The spectra cover the wavelength range from about 3781\AA \;to 6910\AA. More detailed information can be found in Table \ref{tab:SNRanges}.
      
      \begin{table}
        \caption{Available data for the individual nights.}
        \centering\begin{tabular}{c c c c}
          \hline\\[-.5em]
          Night &       Number of               &       Total exposure        &       S/N range\\
                &       Spectra &       [seconds]       &       \\
          \hline\\[-.5em]
          2013-06-08    &       3       &       1360    &       243 - 296   \\
          2013-06-25    &       10      &       5260    &       273 - 351   \\
          2013-08-02    &       2       &       1092    &       145 - 151   \\
          2013-08-04    &       20      &       10000   &       215 - 311   \\
          2013-09-05    &       4       &       2400    &       191 - 248   \\
          2013-09-09    &       13      &       7810    &       122 - 265   \\
          2013-09-30    &       39      &       17100   &       179 - 388   \\
          \hline\\[-.5em]
        \end{tabular}
        \label{tab:SNRanges}
      \end{table}

      Despite the brightness of the target, some cloudy nights decreased the expected S/N (e.g., S/N$\sim$150 after an exposure of 600s on 2013-08-02, while on 2013-09-30 we managed to obtain $\sim$390 after 450s).
      
    \subsection{Data reduction}
      To reduce the data and create the CCFs for each spectrum, the \textit{HARPS DRS} \citep{2003Msngr.114...20M} was used. The data were reduced using the default settings and were then fed to the CCF calculation recipe, used with a weighted $G2$ mask \citep{2002A&A...388..632P}. 
      
      Selecting an optimized CCF computation step was of critical importance. During the detection process, the CCFs need to be shifted in radial velocity (see below), which implies an interpolation between consecutive pixels. The errors in this interpolation can be minimized and its precision increased by selecting the smallest possible step. On the other hand, the computing time increases as the step size decreases. Therefore we settled for a {$\rm 50\; m\; s^{-1}$} step, a good compromise between computing time and high precision. 
      
      The CCF width also requires particular attention because we require a window wide enough to cover the planet's orbital path. Since the expected planet semi-amplitude is about 130 $\rm km\; s^{-1}$ \citep[e.g.,][]{2013ApJ...767...27B}, we selected a window of 175 $\rm km\; s^{-1}$ on each side of the stellar radial velocity (centred on the stellar CCF). This allowed covering the planet's full orbit while leaving on each side of the corresponding RV a continuum section large enough to estimate the noise level.

    \subsection{Calculating the best orbital solution}  \label{sec:Orbital}

      Our initial ephemerides for the orbit of \object{51 Peg b} were taken from \cite{2006ApJ...646..505B}. However, the obtained HARPS data allow us to derive precise radial velocities\footnote{The radial velocities derived with the HARPS DRS pipeline can be found in the online version.} that can be combined with other measurements from the literature, so that we cover a baseline of RV measurements spanning almost 20 years with {different facilities}{(see Table \ref{tab:Measurements})}.

    \begin{table}
          \centering\caption{\object{} Radial velocity data for 51 Peg used to derive the orbital parameters. }
          \centering\resizebox{\columnwidth}{!}{%
            \begin{tabular}{l c c c}
              \hline\\[-.5em]
              Instrument                & Number        & $RV_{\rm System}$ & Reference\\
                                        & of points     & {[$\rm km\; s^{-1}$]}          & \\
              \hline\\
              ELODIE@OHP        &       153             & -33.2516732         & (1)\\
              KECK, AAT, Lick   &       256             & -0.0020280    &(2)\\
              \hline
            \end{tabular}%
          }
          {\tablebib{(1) \citet{2004A&A...414..351N}; (2) \citet{2006ApJ...646..505B}.}\tablefoot{$RV_{\rm system}$ corresponds to the radial velocity of the system as measured by the corresponding instrument.}}
          \label{tab:Measurements}
        \end{table}

      We have thus re-derived the orbital parameters of 51\,Peg\,b. These were computed using the code $YORBIT$ \citep{2011A&A...535A..54S}. This combined dataset allowed us to derive a precise set of orbital parameters for the star and its planetary companion. The derived orbital parameters can be found in Table \ref{tab:51PegYORBIT}, where the value of $RV_{\rm system}$ was set to the HARPS value\footnote{In the fit a different zero point was fitted to each instrument's radial velocities.}. During the fitting process, the eccentricity was fixed to zero because the obtained value was not statistically significant.

        \begin{table}
          \centering\caption{Basic orbital parameters for \object{51 Peg b} as fitted by $YORBIT$}
          \centering\begin{tabular}{l l}
            \hline\\[-.5em]
            Parameters  & Value \\
            \hline\\
            $RV_{\rm system}$ {[$\rm km\; s^{-1}$]}     &       $-33.152$       \\
            Period {[days]}                                             &       $4.231$ \\
            e                                                                   &       $0.0 (fixed)$                \\
            a { [AU]}                                                           &       $0.052$ \\
            $k_{Star}$ {[$\rm m\; s{-1}$]}                                                      &       $55.2$  \\
            $m_2\;sin(i)$ $\rm [M_{\rm Jup}]$                           &       $0.450$ \\
            $\omega$ {[degrees]}                                                &       $0.0 (fixed)$                \\
            $t_{0}$ {[BJD-2400000]}             
&       $56021.256$     \\[.5em]        
            \hline\\[-.5em]
          \end{tabular}
          \label{tab:51PegYORBIT}
        \end{table}

      \subsection{Recovery of the planet signal: methodology}         \label{sec:PlanetRecovery}
        We extracted the planet signal from the stellar noise with the technique described in \cite{2013MNRAS.436.1215M}. In brief, after the CCFs of each observation has been computed, the signal can be recovered in two steps: 
        \begin{itemize}
          \item[step 1:]\label{itm:template} the CCFs are normalized with a stellar template and
          \item[step 2:]\label{itm:stacking} the individual CCFs resulting from step 1 are stacked after correcting for planetary
velocity.\\
        \end{itemize}
        {Especial care in this process was taken for step 1, which consisted in dividing (normalizing) each of the star+planet CCFs by a carefully built stellar CCF template that represents the stellar signal as accurately as
possible. This permitted us to remove the stellar signal so that only the planetary signal and noise were left.} For this template we also needed to ensure the highest possbile S/N so that the division by this template would not introduce significant additional noise. To this end, we constructed two different templates from our observations:
        \begin{description}
          \item[\textit{Template \#1}] This template was constructed by stacking all the CCFs in our sample, after correcting each for the stellar radial velocity induced by the planet. This option implies that the template spectrum also includes a contribution {of the planetary reflected light spectrum}. Despite this fact, since the planet RV varies very rapidly, the planet's minute signal is diluted amidst the noise and can in principle be neglected.\\

          \item[\textit{Template \#2}] This template was constructed by stacking the CCFs of the data collected close to the expected inferior conjunction ephemerids (i.e., the position of the planet in its orbit when its night side faces towards Earth, $\rm 0.9 < \phi < 0.1${, where $\phi$ is the orbital phase}). With this template, the contamination of possible reflected light incoming from the planet is minimized. The downside of this selection is that the number of available spectra for constructing the template is more limited than with template \#1 and therefore might introduce non-negligible noise in the data. For comparison, since only 20 CCFs in our sample are available for template \#2, its S/N is $\sim$40\% of the S/N of template \#1. 
        \end{description}

\begin{figure}
          \includegraphics[width = \columnwidth]{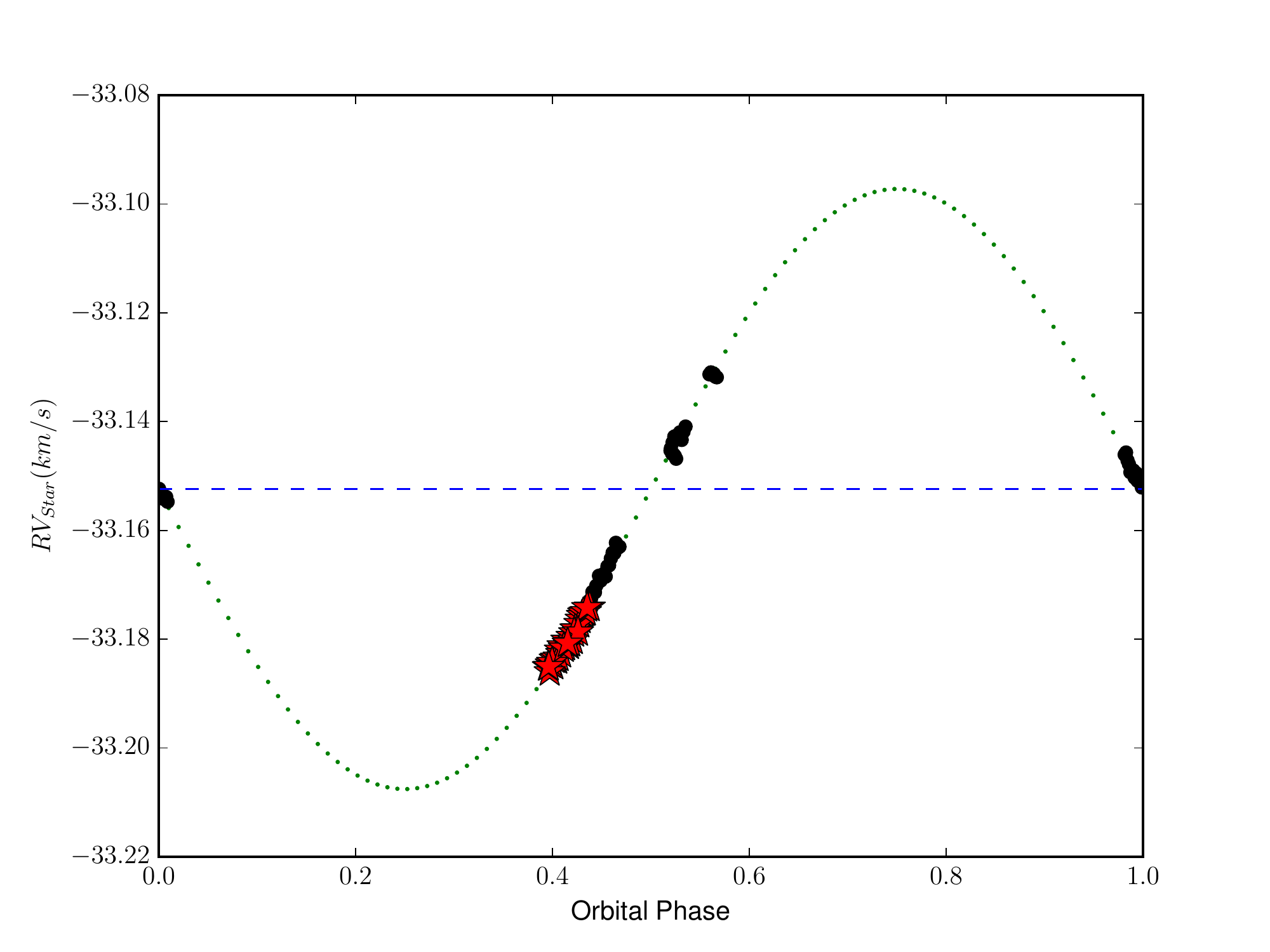}
          \caption{Orbital phases of our data (black circles) and of the observations used to recover the planetary signal (red stars) {at maximum detection significance}. Phase zero corresponds to the inferior conjunction phase, i.e., the position on the orbit where the planetary night side faces Earth. The green dotted line corresponds to the orbital fit from the parameters in Table \ref{tab:51PegYORBIT}.}
          \label{fig:Phases}
        \end{figure}

        To stack the CCFs after normalization by the stellar template (step 2), we discarded observations in which the planetary and stellar signals were spectroscopically blended or close in velocity. To avoid this, we only considered observations (after correcting for the planet's RV computed assuming an elliptical orbit with the orbital parameters in Table \ref{tab:51PegYORBIT}) in which the radial velocity difference between the planet and the star exceeds $\rm 8 \times FWHM_{Star}$ (eight times the FWHM of the stellar CCF). For a more detailed explanation see \citet{2013MNRAS.436.1215M}. Figure \ref{fig:Phases} shows the orbital phases of our data (black circles) and of the observations used to recover the planetary signal (red stars) {at maximum detection significance.}
        
        For each individual CCF, the expected radial velocity of the planet was computed from the Keplerian fit presented in Table \ref{tab:51PegYORBIT}. The different processed CCFs were then re-centered by subtracting the planetary radial velocity from all of them (i.e., centering them around the velocity of the planet at any given moment) and co-adding them to increase their S/N. 
        
        After the individual CCFs were stacked, a Gaussian curve was fitted and its amplitude was used to compute the detection significance. To discard fits with no physical meaning, {two} constrains were set in place:
        \begin{itemize}
          \item $\rm FWHM_{Planet} > 0.9\; FWHM_{Star}$ - {This lower limit takes into account noise that might decrease the width of the planet's CCF. For instance, if the star's convective envelope were to be tidally locked to its planetary companion, the planet would "see" its host star unaffected by the star's rotation and therefore the planetary CCF might be an unbroadened version of the star's \citep[for further details we refer to][]{1999ApJ...522L.145C}. This is clearly not the case for the 51 Peg system, as the stellar rotation period \citep[$\sim$ 21 days, see][]{2010MNRAS.408.1666S} is much longer than the planetary orbital period ($\sim$ 4 days), and therefore this effect should be negligible.}
          \item $\rm FWHM_{Planet} < 4\; FWHM_{Star} $ - Planetary rotation might cause the planetary CCF to be broadened relatively to the CCF of the host star while allowing for enough of a continuum for noise estimation. This limit was computed from the extreme case of a planet with twice the size of Jupiter but the same rotation period and observed edge-on. Nonetheless, we expect the rotation period of a close-in giant planet to be much lower due to tidal interaction
with its star. 
        \end{itemize}
        
        The significance of this detection $D$ is then defined by 
        \begin{equation}
          D = \left|\frac{A}{\rm \sigma_{noise}}\right|
          \label{eq:DetectSig}
        ,\end{equation}
        where $A$ is the amplitude of a Gaussian fit to the planet's signal and $\rm \sigma_{noise}$ is the continuum noise on both sides of the signal. {We define the continuum noise as the standard deviation of the pixel intensity of the stacked CCFs at a separation of more than $\rm 2\times FWHM_{Planet}$ from the center of the detected signal.} 
        
        As mentioned above, the planet's real mass, and therefore its real orbital velocity semi-amplitude $k_{\rm Planet}$, is not known. To pinpoint this real semi-amplitude, we computed the detection significance over a range of planetary orbital semi-amplitudes ({75-275 $\rm km\; s^{-1}$}) centered on the semi-amplitude ($k_{\rm Planet}\sim$ 133 $\rm km\; s^{-1}$) calculated from the minimum mass recovered from the radial velocity fit presented in Table \ref{tab:51PegYORBIT}.
        
    \subsection{Characterizing the planet}
      When stacking the CCFs, we can expect that the maximum detection significance occurs when the correct velocity shift is used in each individual CCF that we combine. In particular, this is expected to occur if we are able to input the correct semi-amplitude radial velocity signal of the planet. A significant detection of the planetary signal (and its radial velocity semi-amplitude) will then allow inferring the planet-star mass ratio $q$ from
      \begin{equation}
         q \equiv \frac{M_{\rm Planet}}{M_{\rm Star}} =  \frac{k_{\rm Star}}{k_{\rm Planet}}
        \label{eq:MassRatio}
      ,\end{equation}
      where $k_{\rm Planet}$ is the most significant amplitude as delivered from the analysis of the previous section, and $k_{\rm Star}$ as derived in Sect. \ref{sec:Orbital}. With $q$, and given the derived value for the stellar mass ($1.04\,M_\odot$), we can compute the real mass for the planet. Together with the minimum mass, this value allows deriving the planetary orbital inclination.

    \begin{figure*}
        \hfill\includegraphics[width = \columnwidth, page =1 ]{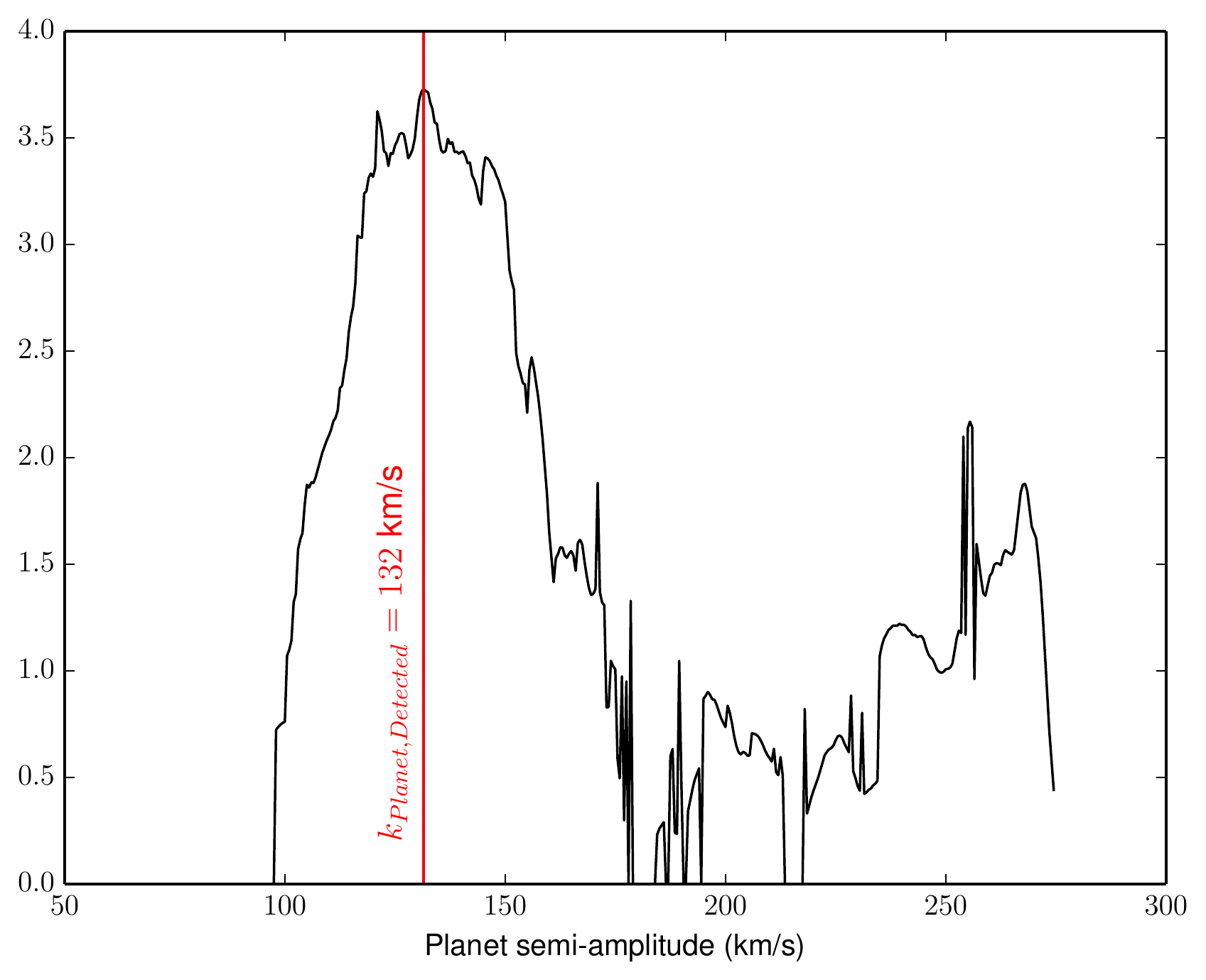}\hfill
        \includegraphics[width = \columnwidth, page =1 ]{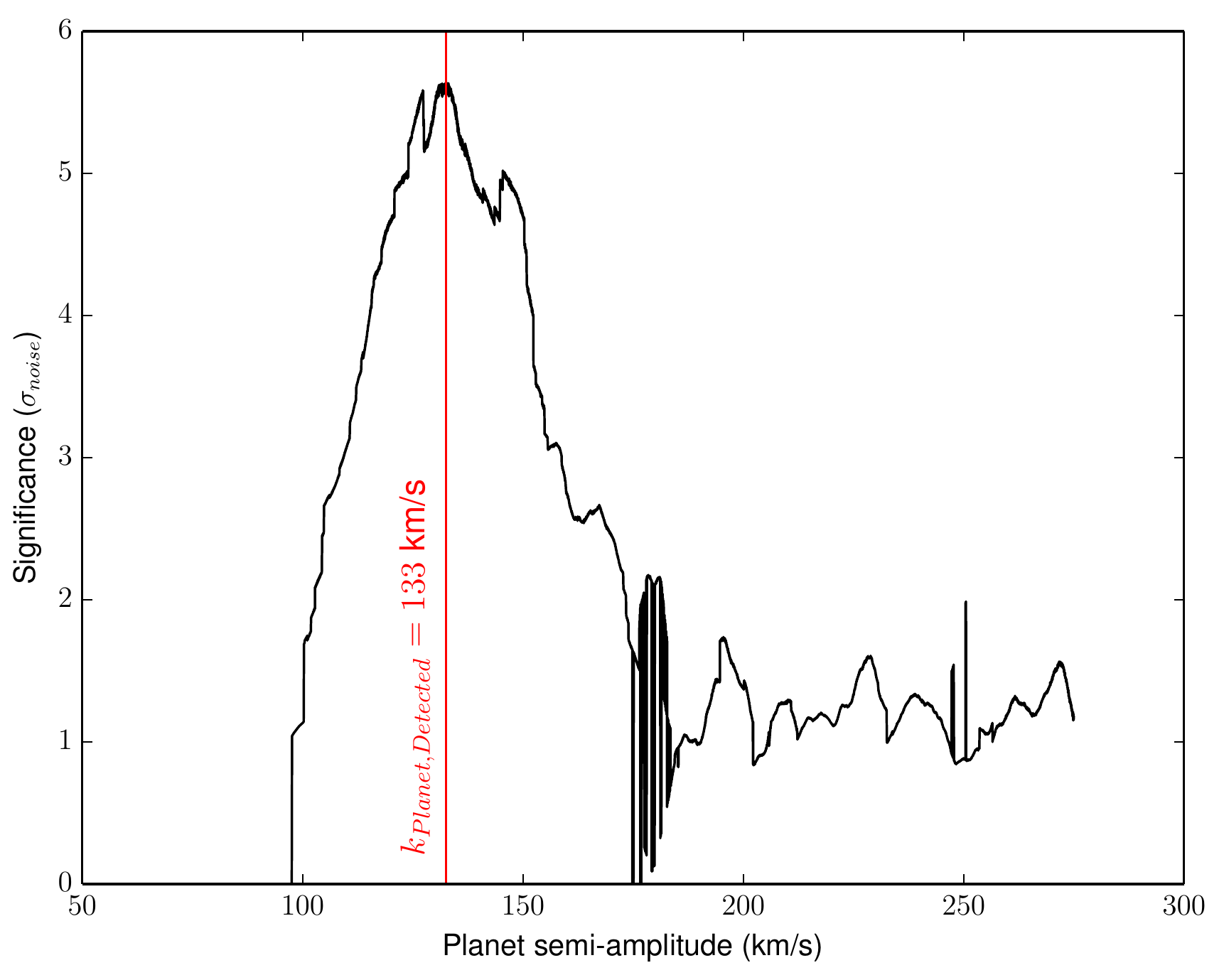}\hfill
      \caption{Detection significance as a function of $k_{\rm Planet}$. The red line corresponds to the $k_{\rm Planet}$ value for maximum detection. The maximum detection occurs for similar $k_{\rm Planet}$ values for both templates. The amplitude values set to zero corresponds to $k_{\rm Planet}$ values for which no Gaussian fit with our restrictions could be achieved. \textit{Left panel:} using template \#1; \textit{right panel:} using template \#2.}
      \label{fig:AmpK2}
    \end{figure*}

    \section{Results}                                           \label{sec:Results}

     \begin{figure} 
        \includegraphics[width = \columnwidth, page =2 ]{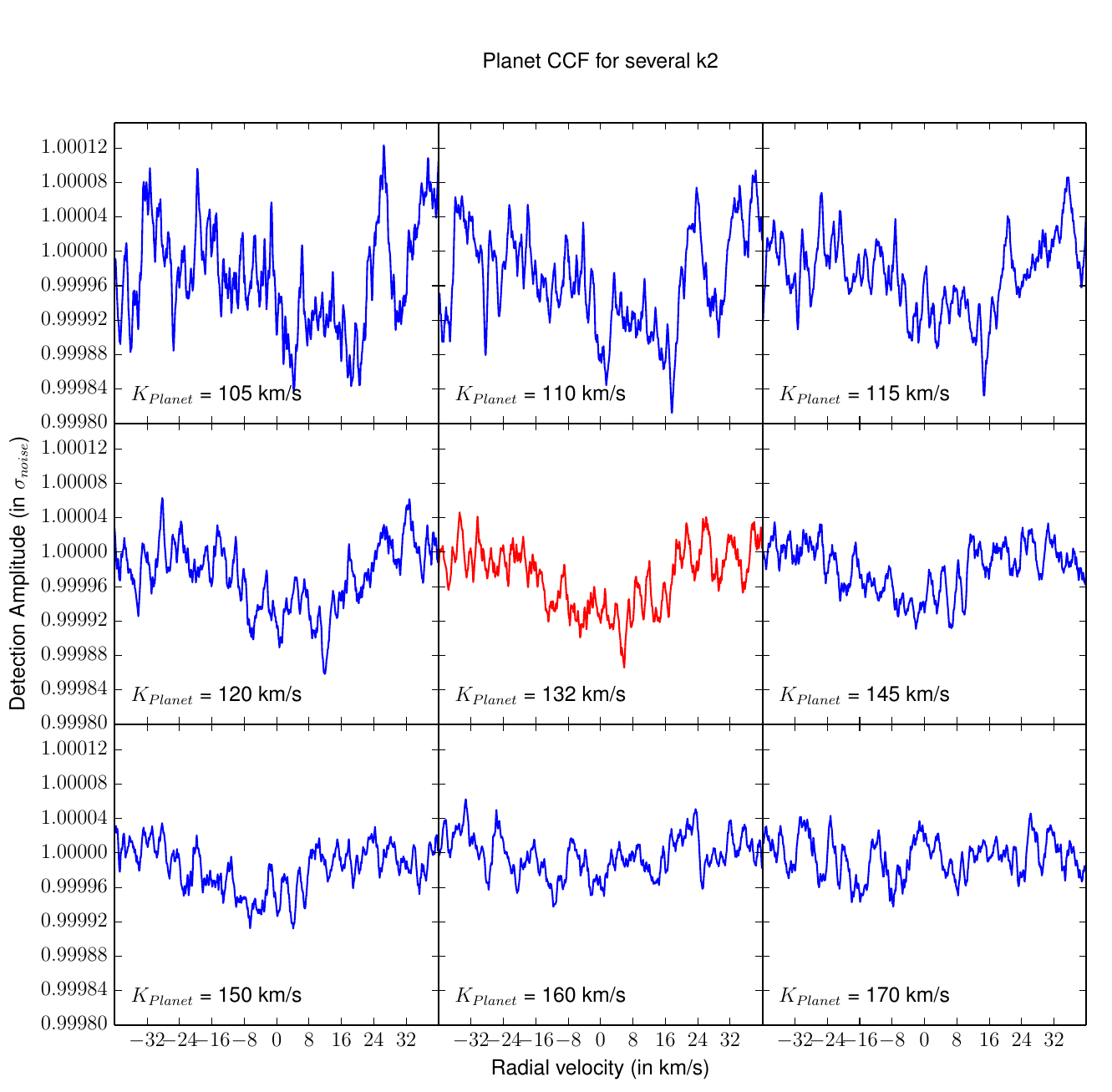}
      \caption{Detected signals as a function of $k_{\rm Planet}$ over the selected velocity range. With decreasing distance to the maximum detection, the signal becomes better defined and the continuum noise less dispersed.}
      \label{fig:DetectionGRid}
    \end{figure}

    Following the method described in Sect. \ref{sec:Method}, we calculated the detection significance {for evenly distributed radial velocity semi-amplitudes of the planet {75 $\rm km\; s^{-1}$ < $k_{\rm Planet}$ < 275 $\rm km\; s^{-1}$},, with a step of 0.05 $\rm km\; s^{-1}$}. For each $k_{\rm Planet}$ we stacked the individual CCFs after correcting for the corresponding radial velocity\footnote{{The constraints mentioned in Sect. \ref{sec:PlanetRecovery}
cause the number of spectra used for constructing each combined CCF to vary with the assumed value of $k_{\rm Planet}$ (it increases with $k_{\rm Planet}$). However, the total S/N of the resulting CCF is always above the S/N threshold for detection of the observed signal (with an amplitude of $\rm 6\pm0.4 \times 10^{-5}$).}}. {We then performed a simple Gaussian fit to the resulting stacked CCF, using the restrictions discussed in Sect. \ref{sec:Method}. }The significance of the detection (D) was then derived using Eq. \ref{eq:DetectSig}. This process was repeated for the two different template options presented in Sect. \ref{sec:Method}.   
     
    {Template \#1 yields a maximum detection significance of 3.7-$\rm \sigma_{noise}$ for {$k_{\rm Planet}$ = 132 $\rm km\; s^{-1}$}, while with Template \#2 a maximum significance of 5.6-$\rm \sigma_{noise}$ is computed for {$k_{\rm Planet}$ = 133 $\rm km\; s^{-1}$}. To be conservative, the error bars on the value derived using template \#2 were computed from the 2-$\sigma$ uncertainty in the amplitude of the Gaussian fitted to the CCF of the planet at maximum detection significance, yielding $k_{\rm Planet} = 133^{+19}_{-20}$ {$\rm km\; s^{-1}$}. Using the same procedure to compute the error bars for the results using template \#1 would lead to $k_{\rm Planet} = 132^{+2}_{-11}$ {$\rm km\; s^{-1}$}. However, for this latter case, a simple visual inspection of Fig. 2 shows that the significance curve is relatively flat for values between about 120 and 150 $\rm km\; s^{-1}$ (but always above 3-$\rm \sigma_{noise}$ in this range). We thus decided to adopt all the values of $k_{\rm Planet}$ with a significance higher than 3.0-$\rm \sigma_{noise}$, that is, $k_{\rm Planet} = 132^{+19}_{-15}$ {$\rm km\; s^{-1}$}. This of course lowers the significance of our detection to 3.0-$\rm \sigma_{noise}$ (and not 3.7-$\rm \sigma_{noise}$ as derived with the maximum value).}
    
    Following the constraints presented in Sect. \ref{sec:PlanetRecovery}, the planetary CCF for maximum significance was constructed from stacking 25 observations of the available {90} ($\sim$27\%). Although template \#2 yields a higher detection significance, we decided to use template \#1 for the remainder of the paper since it has a much higher S/N because many more CCFs were used to construct it. Therefore it will represent the stellar signal more reliably and is less prone to introduce additional noise into the CCF (including correlated noise that is difficult to quantify).
    
{Table \ref{tab:GaussParams} shows the parameters for the recovered CCF derived from the Gaussian fit using template \#1. In this table, the amplitudes of the star (0.48) and planet ($\rm 6\pm0.4 \times 10^{-5}$) CCFs correspond to the depth of the CCF when the the continuum has been normalized to 1. The planet-to-star flux ratio will then be given by the ratio of these two quantities. The significance of the detection corresponds to the maximum significance for 75 $\rm km\; s^{-1}$ < $k_{\rm Planet}$ < 275 $\rm km\; s^{-1}$ and was computed by dividing the amplitude of the planetary CCF by the noise on the wings of the CCF. The FWHM is the full width at half maximum of the fitted Gaussian, 7.43 $\rm km\; s^{-1}$ for the star, $\rm 22.6 \pm 3.6$ $\rm km\; s^{-1}$ for the planet. }

    Figure \ref{fig:DetectionGRid} shows for several assumed values of $k_{\rm Planet}$ over the adopted range a section of the stacked CCFs centered on the radial velocity  where the planetary signal would be expected from a Keplerian fit to the observations using the parameters in Table \ref{tab:51PegYORBIT} and the selected $k_{\rm Planet}$. {The plot illustrates how the recovered CCF changes when we co-add spectra for different assumed semi-amplitude values. When a value for the semi-amplitude close to 132 $\rm km\; s^{-1}$ is assumed, the recovered signal is well defined and its wings are less noisy; this is closer to the distribution of the expected Gaussian shape of a CCF. However, as the assumed value of the semi-amplitude departs from 132 $\rm km\; s^{-1}$, the CCFs of the {individual} observations will be co-added to each other, albeit imperfectly. Therefore, in these cases a signal is also expected to be detected, but spread out across the {continuum} and of lower significance because the wings additionally contribute to noise. When close enough to the best-fit value of the semi-amplitude, a signal above the noise level can still be seen. This signal might seem to be of similar amplitude as the one detected for the correct value of orbital semi-amplitude, but it will have a lower significance due to the increased noise in the wings. This can be seen in the panels for $k_{\rm Planet}$ = 115 and 120 $\rm km\; s^{-1}$, which show signals that might seem to
be of similar amplitude to the one in the central panel, but because of the increased noise in the wings, these signals are of lower significance according to Eq. \ref{eq:DetectSig} (see Fig. \ref{fig:AmpK2}).} 
     
     \begin{figure} 
        \includegraphics[width = \columnwidth, page =3 ]{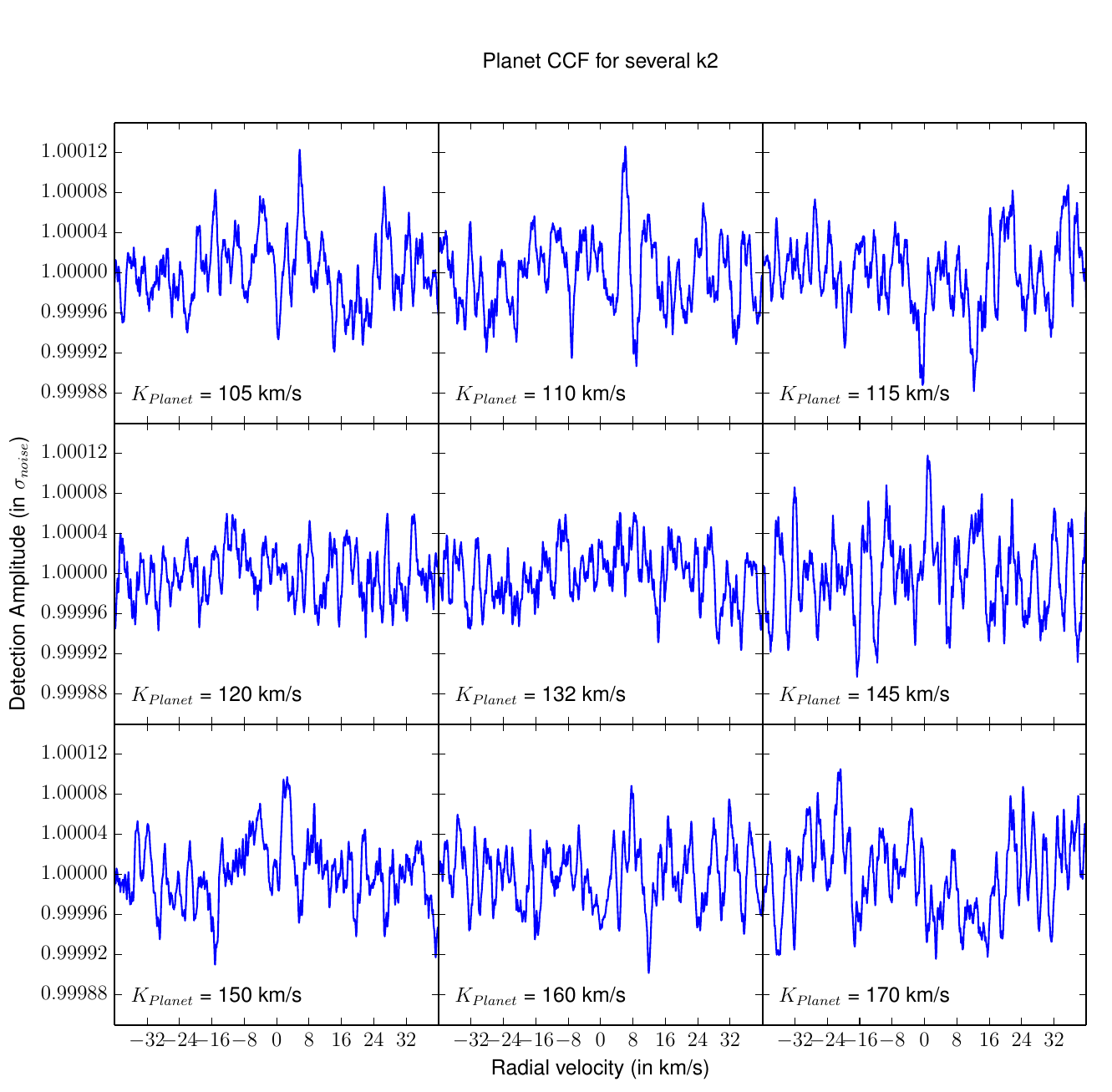}
      \caption{Detected signals as a function of $k_{\rm Planet}$ , but for random sections of the normalized CCF where the planetary signal is not expected to be found. }
      \label{fig:DetectionGRidNull}
    \end{figure}
    
    To test whether the observed signal might be a spurious combination of random noise, we repeated the process, but instead of using a Keplerian function to compute the expected radial velocity signal of the planet, we attributed a random radial velocity within the range [-50, 80] $\rm km\; s^{-1}$ to each
CCF. The result of that analysis can be seen in Fig. \ref{fig:DetectionGRidNull}. They show that no significant signal is detected, as expected. This test gives us confidence that the detected CCF of the planet is not a mere artifact. {The dips that appear in the panels of Fig. \ref{fig:DetectionGRidNull} appear at random positions, and their FWHM is always lower than the FWHM of the star. For instance, the FWHM of the center dip in the 115 $\rm km\; s^{-1}$ panel of Fig. \ref{fig:DetectionGRidNull} has a FWHM of only 3.8 $\rm km\; s^{-1}$. This strongly suggests that these peaks are purely caused by noise, especially as they appear randomly positioned. The constraints that we have specified in Sect. \ref{sec:Method} ensure these dips as discarded as nonphysical and are not confused with the planetary CCF.}
    
    {To verify whether our analysis data-processing was sound, we simulated 100 sets of noiseless CCF functions simulating idealized observations of the  star + planet signal for random values of the planetary semi-amplitude $k_{\rm Planet}$ in the range [100, 180] $\rm km\; s^{-1}$. These sets of simulated observations consisted of {90} star+planet CCFs (the number of observations we have), where the planet radial velocity is computed from one of the randomly selected $k_{\rm Planet}$ through Eq.\,\ref{eq:MassRatio}. Each simulated CCF was built by positioning the stellar Gaussian at the observed velocity of the star at each given moment (obtained from the real CCFs) and co-adding the planet Gaussian with an {amplitude} of $5\times 10^{-5}$ (a value similar to the expected for our test subject) . The goal was to verify whether the injected signal was always fully recovered by the reduction process. The results show that the injected signal was successfully recovered in all simulations. The obtained values of $k_{\rm Planet}$ were always the values that were injected into the expected figure, with a standard deviation of only 0.11\%. Similar results were obtained for the {amplitude}, which was always recovered with a standard deviation lower than 0.01\%. This test shows that that the data-analysis pipeline works correctly and can be used safely to retrieve the planetary signal.}
        
    \begin{figure} 
        \includegraphics[width = \columnwidth, keepaspectratio =true, page = 1]{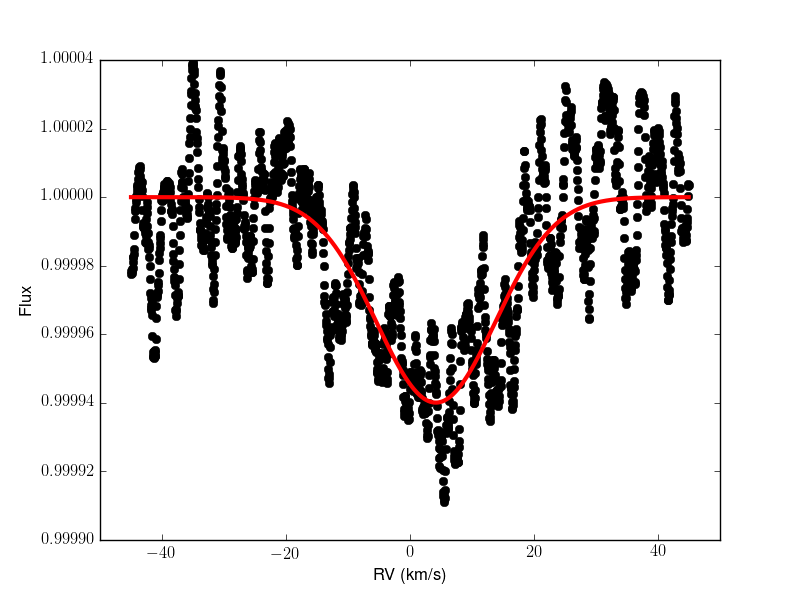}\hfill
      \caption{Planet signal for maximum detection and fitted Gaussian curve with the Levenberg-Marquardt method.}
      \label{fig:PlanetCCF}
    \end{figure}

    The planetary CCF with the highest recovered significance is plotted in Fig. \ref{fig:PlanetCCF} together with a Gaussian fit using a Levenberg-Marquardt algorithm. After removing the planetary CCF and fitting it with a Gaussian, we used the residuals of the fit to define the error bars. To do so, we injected into the residuals a Gaussian curve with the same parameters as the detected planetary CCF after
subtracting the detected Gaussian, but at different radial velocities. This procedure was repeated 10\,000 times, and in each we recovered the signal injected by fitting a Gaussian profile. The standard deviation of the recovered FWHM and {amplitude} values were then considered to be the 1-$\sigma$ errors as listed in Table\,\ref{tab:GaussParams}.

    \begin{table}
      \caption{{Comparison of the stellar and planetary CCF parameters. }}
      \centering\begin{tabular}{l c c l}
        \hline\\[-.5em]  
        Parameter       &       Star    &       {Planet}        &\\
                                &               &                       &\\
        \hline\\
        {Amplitude}     &       0.48    &       6.0$\pm$0.4 &\hspace{-1.em}$\rm \times$ $10^{-5}$\\
        Significance $[\rm \sigma_{noise}]$ &--&        3.7$\pm$0.2&\\
        FWHM {[$\rm km\; s^{-1}$]}      &       7.43    &       \hspace{-.5em}22.6$\pm$3.6&\\
        \hline\\
      \end{tabular}
      {\tablefoot{Comparison of the stellar and planetary CCF parameters. For the star signal, we present the median value of the amplitude and FWHM of its CCF for all observations. For the planetary CCF, we present the values of the amplitude and FWHM of its Gaussian fit and its detection significance. In both cases, the level of the CCF continuum flux has been set to one.}}
      \label{tab:GaussParams} 
    \end{table}

  \section{Discussion}                                          \label{sec:Discussion}
    {Our results suggest that we were able to successfully detect the light spectrum of 51\,Peg reflected on its hot Jupiter companion. The results also indicate that the best-fit semi-amplitude of the orbital motion of the planet is $k_{\rm Planet} = 132^{+19}_{-15}$ $\rm km\; s^{-1}$.}.
        
    {The planetary mass cannot be lower than its real mass (corresponding to an orbital inclination of 90 $^{\circ}$), which places a higher limit on the planetary orbital semi-amplitude of $k_{\rm Planet} \sim 133$ $\rm km\; s^{-1}$. Combining the detected value of the semi-amplitude with this assumption, Eq. \ref{eq:MassRatio} yields a real mass for \object{51 Peg b} of $0.46^{+0.06}_{-0.01}\;\rm M_{Jup}$ for a stellar mass of $\rm 1.04\; M_{\odot}$ \citep{2013A&A...556A.150S}. By comparing this with the derived minimum mass of $m_2\;sin\;i = 0.45 \;\rm M_{Jup}$ (Table \ref{tab:51PegYORBIT}), we can infer an orbital inclination of $80^{+10}_{-19}$ degrees. This result is compatible with the results obtained independently
by \citet{2013ApJ...767...27B} - $79.6^{\circ}<I<82.2^{\circ}$  for a $1\;\sigma$ confidence level. }
    
    {These results suggest that we were able to successfully recover the reflected planetary signature, but the noise present in the data hinders constraining parameters like the planet's geometric albedo. It is interesting, however, to verify which possible values of the albedo we would recover if we assumed plausible  values for the radii of the planet according to Eq. \ref{eq:FluxesRatio} and assumed that the planet-star flux ratio is given by the ratio of the amplitudes of the the detected planet and stellar CCFs. Note that the scatter of radii observed for hot-Jupiter planets of a given mass does not allow us to simply use the derived planetary mass to estimate a reliable value for the planetary radius \cite[e.g.,][]{2010RPPh...73a6901B, 2010SSRv..152..423F}.}
    
    {The results show that the obtained $A_{\rm g}$ is above unity for a radius of 1.2 $\rm R_{Jup}$ . Furthermore, if the planet's reflectivity is assumed to be Lambertian (i.e., the planetary surface reflects isotropically in all directions), it has been shown that the planetary albedo will have at most
a value of $\text{}\text{two thirds}$ \citep{1975lpsa.book.....S}. Higher values imply that the planetary surface or atmosphere would be strongly backscattering. It has been shown that cloudy planets in the solar system tend to be close to Lambertian \citep[e.g., Venus - see ][]{2011exop.book..111T}, while rocky planets tend to be more uniformly bright than a Lambert's sphere \citep[e.g., Earth and Moon - see][]{2011exop.book..111T}. Moreover, literature values suggest that hot Jupiters usually have albedos lower than $\sim$0.3  \citep[][]{2011ApJ...729...54C}. However, several works \citep[e.g., the case of HD 189733b,][]{2013ApJ...772L..16E,2013MNRAS.432.2917P} reported albedo values as high as 0.4-0.5. Such high values may be explained by the scattering of condensates in the high atmosphere \citep{2003ApJ...588.1121S}. }

{For our measurements, the observations used to recover the planetary signal were obtained at an average orbital phase $g(\alpha)=0.87$. Thus, for a perfectly reflecting Lambert sphere, that is, $A_{\rm g} = 2/3$, Eq. \ref{eq:FluxesRatio} yields a radius for the planet of 1.6$\pm$0.2 $\rm R_{Jup}$ (the error bar denotes the error found in the detected CCF amplitude). For an albedo of 0.5, the planetary radius will increase to 1.9$\pm$0.3 $\rm R_{Jup}$. Such high radii have been observed in other hot Jupiters of similar mass and orbital period as 51 Peg\,b \citep[e.g., Kepler-12\,b,][]{2011ApJS..197....9F}, making our assumptions plausible.}

    The estimates discussed here are based on a {3-sigma detection, however}. As we can see in Figure\,\ref{fig:PlanetCCF}, the detection we obtained shows a strongly correlated noise signature, which is difficult to take into account when fitting the observed signal with a Gaussian function. We can thus not fully trust the derived parameters for the detected planetary CCF. This is crucial for the FWHM of the recovered CCF, which might be used to estimate the planetary rotation velocity \citep[e.g.,][]{2002A&A...392..215S}. Such a high CCF width, when compared with the stellar CCF, would imply an extremely rapid rotation for the planet ($\sim$18 $\rm km\; s^{-1}$), much higher than expected for a planet assumed to be tidally locked to its host star ($\sim$ 2 $\rm km\; s^{-1}$). Nonetheless, because of our low confidence in the recovered CCF parameters, any conclusion regarding the planetary rotation velocity would be too speculative. 
    
    An alternative explanation for the broadening might be an imperfect orbital solution. But our target has been followed for many years, and the star's orbital parameters are very well constrained. Furthermore, the errors in the stellar RV are on the order of the $\rm m\; s^{-1}$, which transposed into the planet RV domain would yield at most errors in RV of {2-3} $\rm km\; s^{-1}$. This is insufficient to explain such broadening.
    
    {Even {though there is no stellar (visual) companion to 51 Peg reported in the literature }\citep[e.g., see ][]{2002ApJ...566.1132L}, we explored the possibility that the observed signal might be caused by the spectrum of a contaminating background star. We used the Besan\c{c}on model \citep{2003A&A...409..523R} to determine the probability of a stellar contaminant close to 51 Peg, whose spectrum would cause the magnitude of the observed signal. Given the area of the CCF for the (here suggested) planetary signal ($10^{-4}$) and the magnitude of 51 Peg, the contaminant should be of magnitude 16 or brighter (brighter if the contaminant does not fall entirely within the HARPS fibre, which has a diameter of 1 arcsec). According to the Besan\c{c}on model, there are only 760 FGK and M stars per square degree at the position of \object{51 Peg b} that possess a magnitude brighter than this value\footnote{We only chose FGK and M stars since other spectral types would not produce a CCF detectable by the HARPS reduction, which used a G2 template for the cross-correlation.}  . If we conservatively assume that any contaminant at a distance smaller than 2.6 arcsec were able to produce the observed signal \citep[for details see][]{2013A&A...550A..75C}, we can expect a probability of a stellar contaminant of at most only 0.1\%. We can thus reasonably confidently discard the possibility that the signal is produced by a background stellar contaminant. }
    
    {Note that the sky brightness has a magnitude per arcsec$^2$, in V band, of 21.8 on a new moon night, and of 20.0 on a full moon night \citep[e.g., see ][]{2013A&A...550A..75C}, which excludes this as a possible source of contamination.}
    
    {A small offset can also be seen on the center of the CCF
recovered with the highest detection significance, which can be expected to be centered around zero for the real $k_{\rm Planet}$. Several explanations can be found for this offset: 
      \begin{inparaenum}[i)]
        \item it might be physically induced, for instance by high-speed winds from the day to the night side in the planetary atmosphere \citep[ see ][]{2012ApJ...751..117M};
        \item it might be an observational bias that results from stacking observations of the planet all on the same side of the superior conjunction while not at full phase (see Fig. \ref{fig:Phases});
        \item and it might be a higher contribution of noise in one of the sides of the CCF.
      \end{inparaenum}
    However, we consider that this offset is not significant because of the high noise.}
    
\begin{figure}
     \includegraphics[width = \columnwidth]{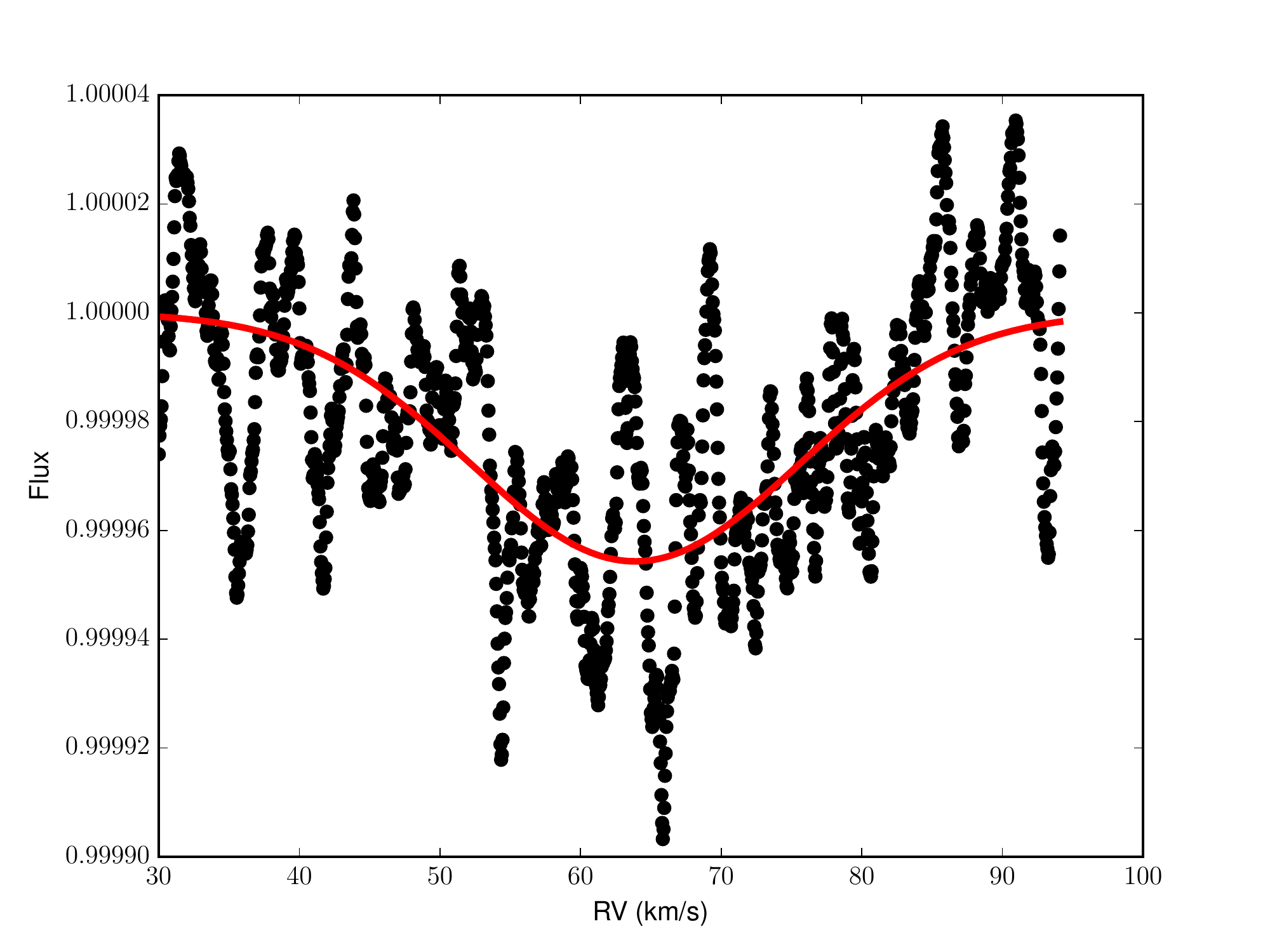} 
     \caption{Simulated signal injected at +60\,km\,s$^{-1}$ in relation to the position of the center of the planetary CCF.}
     \label{fig:SimulatedSignal}
    \end{figure}

    To test how much the observed signal can be affected by this noise, we performed the following test. For each observed CCF we injected an artificial signal (a pure Gaussian) with an amplitude of $5\times10^{-5}$ (similar to the retrieved planetary signal), but at a different position in velocity (at +60\,km\,s$^{-1}$ when compared to the planet). We then ran our data-reduction process and tried to retrieve this signal. As can be seen in Fig.\,\ref{fig:SimulatedSignal}, we recovered the injected CCF with a similar amplitude of $4.6\times10^{-5}$. However, the FWHM of the fitted Gaussian is $\sim\;$27.6\,km\,s$^{-1}$, significantly larger than the injected FWHM value (7.43\,km\,s$^{-1}$ as observed in the stellar CCF). This test was repeated for different signal intensities, and we found that when the signal strength is much higher than the noise, it is recovered with parameters close to the injected values. Nonetheless, when the signal strength is close to the noise level, the recovered signal appears broadened. This test shows that the parameters of the recovered CCF are strongly affected by the noise.
    
In this particular simulation we show that the obtained FWHM is much broader than the injected one. Interestingly, the value for the FWHM we recovered in our data for the planetary signal (Table\,\ref{tab:GaussParams}) is also much higher than the value observed for the stellar CCF. This shows that the FWHM we obtain cannot be considered reliable to derive the rotational velocity of the planet or the velocity of its atmospheric winds, for instance \citep[e.g.,][]{2010Natur.465.1049S}.

  \section{Conclusions}                                         \label{sec:Conclusions} 
  
  We presented the first application of the technique described by \cite{2013MNRAS.436.1215M}. We were able to {find evidence for the reflected light from} the hot-Jupiter 51 Peg\,b with
this technique. {This result is encouraging} and constitutes a very valuable proof of concept. Our method can be used to recover an exoplanet's spectroscopic reflected signature from among the stellar noise, despite the extremely low planet-to-star flux ratio.
  
  {Although only a 3 $\rm \sigma_{noise}$ detection was possible, the obtained data allow us to place constraints on the planetary mass ($\rm 0.46^{+0.06}_{-0.01}\;M_{Jup}$) and orbital inclination ($\rm 80^{+10}_{-19}$ degrees), assuming
the detection is real. As presented above, reasonable values for the albedo (0.5) are found if we assume that \object{51 Peg b} is an inflated hot Jupiter with a radius of about $1.9\;\rm R_{Jup}$. Strongly inflated planets with similar mass and orbital period have been reported in the literature (with radii measured with transits).}
    
  Of particular interest is the fact that this detection was possible with data collected with a currently existing observing facility on a 4-meter-class telescope (the HARPS spectrometer at the 3.6 m ESO telescope). Additional data will also allow additionally constraining of some of the inferred parameters by increasing the detection significance and allowing for a better fit to the detected planetary CCF. 
  
  These encouraging results clearly show a bright future for this type of studies when next-generation instruments (e.g., ESPRESSO at the VLT) and telescopes (e.g., ESO's E-ELT) become available to the community. The sheer increase in precision and collecting power will allow for the detection of reflected light from smaller planets, planets on orbits with longer periods, or an increase in detail for larger planets like \object{51 Peg b}.

\begin{acknowledgements}
  This work was supported by the European Research Council/European Community under the FP7 through Starting Grant agreement number 239953. P.F. and N.C.S. acknowledge support by  Funda\c{c}\~ao para a Ci\^encia e a Tecnologia (FCT) through Investigador FCT contracts of reference IF/01037/2013 and IF/00169/2012, respectively, and POPH/FSE (EC) by FEDER funding through the program ``Programa Operacional de Factores de Competitividade - COMPETE''. S.G.S, acknowledges the support from FCT in the form of the fellowships SFRH/BPD/47611/2008. {We further acknowledge the support from Funda\c{c}\~ao para a Ci\^encia e a Tecnologia (Portugal) through FEDER funds in program COMPETE, as well as through national funds, in the form of the grants with references RECI/FIS-AST/0176/2012 (FCOMP-01-0124-FEDER-027493), RECI/FIS-AST/0163/2012 (FCOMP-01-0124-FEDER-027492), and UID/FIS/04434/2013}. {We acknowledge the referee for the several and interesting points raised and for drawing our attention to important shortcomings of our analysis.}
\end{acknowledgements}

\bibliographystyle{aa}
\bibliography{JorgesBibtexRefs}

\begin{thebibliography}{62}
\expandafter\ifx\csname natexlab\endcsname\relax\def\natexlab#1{#1}\fi

\bibitem[{Alonso {et~al.}(2009)Alonso, Guillot, Mazeh, Aigrain, Alapini, Barge,
  Hatzes, \& Pont}]{2009A&A...501L..23A-eprintv1}
Alonso, R., Guillot, T., Mazeh, T., {et~al.} 2009, Astronomy and Astrophysics,
  501, L23

\bibitem[{{Baraffe} {et~al.}(2010){Baraffe}, {Chabrier}, \&
  {Barman}}]{2010RPPh...73a6901B}
{Baraffe}, I., {Chabrier}, G., \& {Barman}, T. 2010, Reports on Progress in
  Physics, 73, 016901

\bibitem[{Baranne {et~al.}(1996)Baranne, Queloz, Mayor, Adrianzyk, Knispel,
  Kohler, Lacroix, Meunier, Rimbaud, \& Vin}]{1996A&AS..119..373B}
Baranne, A., Queloz, D., Mayor, M., {et~al.} 1996, A\&As, 119, 373

\bibitem[{{Barstow} {et~al.}(2014){Barstow}, {Aigrain}, {Irwin}, {Hackler},
  {Fletcher}, {Lee}, \& {Gibson}}]{2014arXiv1403.6664B}
{Barstow}, J.~K., {Aigrain}, S., {Irwin}, P.~G.~J., {et~al.} 2014, ArXiv
  e-prints

\bibitem[{{Birkby} {et~al.}(2013){Birkby}, {de Kok}, {Brogi}, {de Mooij},
  {Schwarz}, {Albrecht}, \& {Snellen}}]{2013MNRAS.436L..35B}
{Birkby}, J.~L., {de Kok}, R.~J., {Brogi}, M., {et~al.} 2013, \mnras, 436, L35

\bibitem[{{Bonfils} {et~al.}(2013){Bonfils}, {Delfosse}, {Udry}, {Forveille},
  {Mayor}, {Perrier}, {Bouchy}, {Gillon}, {Lovis}, {Pepe}, {Queloz}, {Santos},
  {S{\'e}gransan}, \& {Bertaux}}]{2013A&A...549A.109B}
{Bonfils}, X., {Delfosse}, X., {Udry}, S., {et~al.} 2013, \aap, 549, A109

\bibitem[{Borucki {et~al.}(2013)Borucki, Agol, Fressin, Kaltenegger, Rowe,
  Isaacson, Fischer, Batalha, Lissauer, Marcy, Fabrycky, {D{'e}sert}, Bryson,
  Barclay, Bastien, Boss, Brugamyer, Buchhave, Burke, Caldwell, Carter,
  Charbonneau, Crepp, {Christensen-Dalsgaard}, Christiansen, Ciardi, Cochran,
  DeVore, Doyle, Dupree, Endl, Everett, Ford, Fortney, Gautier, Geary, Gould,
  Haas, Henze, Howard, Howell, Huber, Jenkins, Kjeldsen, Kolbl, Kolodziejczak,
  Latham, Lee, Lopez, Mullally, Orosz, Prsa, Quintana, {Sanchis-Ojeda},
  Sasselov, Seader, Shporer, Steffen, Still, Tenenbaum, Thompson, Torres,
  Twicken, Welsh, \& Winn}]{2013Sci...340..587B}
Borucki, W.~J., Agol, E., Fressin, F., {et~al.} 2013, Science, 340, 587

\bibitem[{{Borucki} {et~al.}(2010){Borucki}, {Koch}, {Basri}, {Batalha},
  {Brown}, {Caldwell}, {Caldwell}, {Christensen-Dalsgaard}, {Cochran},
  {DeVore}, {Dunham}, {Dupree}, {Gautier}, {Geary}, {Gilliland}, {Gould},
  {Howell}, {Jenkins}, {Kondo}, {Latham}, {Marcy}, {Meibom}, {Kjeldsen},
  {Lissauer}, {Monet}, {Morrison}, {Sasselov}, {Tarter}, {Boss}, {Brownlee},
  {Owen}, {Buzasi}, {Charbonneau}, {Doyle}, {Fortney}, {Ford}, {Holman},
  {Seager}, {Steffen}, {Welsh}, {Rowe}, {Anderson}, {Buchhave}, {Ciardi},
  {Walkowicz}, {Sherry}, {Horch}, {Isaacson}, {Everett}, {Fischer}, {Torres},
  {Johnson}, {Endl}, {MacQueen}, {Bryson}, {Dotson}, {Haas}, {Kolodziejczak},
  {Van Cleve}, {Chandrasekaran}, {Twicken}, {Quintana}, {Clarke}, {Allen},
  {Li}, {Wu}, {Tenenbaum}, {Verner}, {Bruhweiler}, {Barnes}, \&
  {Prsa}}]{2010Sci...327..977B}
{Borucki}, W.~J., {Koch}, D., {Basri}, G., {et~al.} 2010, Science, 327, 977

\bibitem[{{Brogi} {et~al.}(2014){Brogi}, {de Kok}, {Birkby}, {Schwarz}, \&
  {Snellen}}]{2014arXiv1404.3769B}
{Brogi}, M., {de Kok}, R.~J., {Birkby}, J.~L., {Schwarz}, H., \& {Snellen},
  I.~A.~G. 2014, ArXiv e-prints

\bibitem[{Brogi {et~al.}(2012)Brogi, Snellen, {de Kok}, Albrecht, Birkby, \&
  {de Mooij}}]{2012Natur.486..502B}
Brogi, M., Snellen, I.~A.~G., {de Kok}, R.~J., {et~al.} 2012, \nat, 486, 502

\bibitem[{Brogi {et~al.}(2013)Brogi, Snellen, {de Kok}, Albrecht, Birkby, \&
  {de Mooij}}]{2013ApJ...767...27B}
Brogi, M., Snellen, I.~A.~G., {de Kok}, R.~J., {et~al.} 2013, \apj, 767, 27

\bibitem[{{Butler} {et~al.}(2006){Butler}, {Wright}, {Marcy}, {Fischer},
  {Vogt}, {Tinney}, {Jones}, {Carter}, {Johnson}, {McCarthy}, \&
  {Penny}}]{2006ApJ...646..505B}
{Butler}, R.~P., {Wright}, J.~T., {Marcy}, G.~W., {et~al.} 2006, \apj, 646, 505

\bibitem[{Charbonneau {et~al.}(2002)Charbonneau, Brown, Noyes, \&
  Gilliland}]{2002ApJ...568..377C}
Charbonneau, D., Brown, T.~M., Noyes, R.~W., \& Gilliland, R.~L. 2002, \apj,
  568, 377

\bibitem[{Charbonneau {et~al.}(1999)Charbonneau, Noyes, Korzennik, Nisenson,
  Jha, Vogt, \& Kibrick}]{1999ApJ...522L.145C}
Charbonneau, D., Noyes, R.~W., Korzennik, S.~G., {et~al.} 1999, \apj, 522, L145

\bibitem[{{Collier Cameron} {et~al.}(1999){Collier Cameron}, Horne, Penny, \&
  James}]{1999Natur.402..751C}
{Collier Cameron}, A., Horne, K., Penny, A., \& James, D. 1999, \nat, 402, 751

\bibitem[{{Collier Cameron} {et~al.}(2002){Collier Cameron}, Horne, Penny, \&
  Leigh}]{2002MNRAS.330..187C}
{Collier Cameron}, A., Horne, K., Penny, A., \& Leigh, C. 2002, \mnras, 330,
  187

\bibitem[{Cowan \& Agol(2011)}]{2011ApJ...729...54C}
Cowan, N.~B. \& Agol, E. 2011, \apj, 729, 54

\bibitem[{{Cunha} {et~al.}(2013){Cunha}, {Figueira}, {Santos}, {Lovis}, \&
  {Bou{\'e}}}]{2013A&A...550A..75C}
{Cunha}, D., {Figueira}, P., {Santos}, N.~C., {Lovis}, C., \& {Bou{\'e}}, G.
  2013, \aap, 550, A75

\bibitem[{{de Kok} {et~al.}(2014){de Kok}, {Birkby}, {Brogi}, {Schwarz},
  {Albrecht}, {de Mooij}, \& {Snellen}}]{2014A&A...561A.150D}
{de Kok}, R.~J., {Birkby}, J., {Brogi}, M., {et~al.} 2014, \aap, 561, A150

\bibitem[{{Deming} {et~al.}(2005){Deming}, {Seager}, {Richardson}, \&
  {Harrington}}]{2005Natur.434..740D}
{Deming}, D., {Seager}, S., {Richardson}, L.~J., \& {Harrington}, J. 2005,
  \nat, 434, 740

\bibitem[{{Demory}(2014)}]{2014arXiv1405.3798D}
{Demory}, B.-O. 2014, ArXiv e-prints

\bibitem[{{Evans} {et~al.}(2013){Evans}, {Pont}, {Sing}, {Aigrain}, {Barstow},
  {D{\'e}sert}, {Gibson}, {Heng}, {Knutson}, \& {Lecavelier des
  Etangs}}]{2013ApJ...772L..16E}
{Evans}, T.~M., {Pont}, F., {Sing}, D.~K., {et~al.} 2013, \apjl, 772, L16

\bibitem[{{Fortney} {et~al.}(2011){Fortney}, {Demory}, {D{\'e}sert}, {Rowe},
  {Marcy}, {Isaacson}, {Buchhave}, {Ciardi}, {Gautier}, {Batalha}, {Caldwell},
  {Bryson}, {Nutzman}, {Jenkins}, {Howard}, {Charbonneau}, {Knutson}, {Howell},
  {Everett}, {Fressin}, {Deming}, {Borucki}, {Brown}, {Ford}, {Gilliland},
  {Latham}, {Miller}, {Seager}, {Fischer}, {Koch}, {Lissauer}, {Haas}, {Still},
  {Lucas}, {Gillon}, {Christiansen}, \& {Geary}}]{2011ApJS..197....9F}
{Fortney}, J.~J., {Demory}, B.-O., {D{\'e}sert}, J.-M., {et~al.} 2011, \apjs,
  197, 9

\bibitem[{{Fortney} \& {Nettelmann}(2010)}]{2010SSRv..152..423F}
{Fortney}, J.~J. \& {Nettelmann}, N. 2010, \ssr, 152, 423

\bibitem[{{Howard} {et~al.}(2010){Howard}, {Marcy}, {Johnson}, {Fischer},
  {Wright}, {Isaacson}, {Valenti}, {Anderson}, {Lin}, \&
  {Ida}}]{2010Sci...330..653H}
{Howard}, A.~W., {Marcy}, G.~W., {Johnson}, J.~A., {et~al.} 2010, Science, 330,
  653

\bibitem[{{Kane} {et~al.}(2011){Kane}, {Gelino}, {Ciardi}, {Dragomir}, \& {von
  Braun}}]{2011ApJ...740...61K}
{Kane}, S.~R., {Gelino}, D.~M., {Ciardi}, D.~R., {Dragomir}, D., \& {von
  Braun}, K. 2011, \apj, 740, 61

\bibitem[{Knutson {et~al.}(2009)Knutson, Charbonneau, Cowan, Fortney, Showman,
  Agol, \& Henry}]{2009ApJ...703..769K-eprintv1}
Knutson, H., Charbonneau, D., Cowan, N., {et~al.} 2009, ArXiv e-prints

\bibitem[{{Knutson} {et~al.}(2014){Knutson}, {Dragomir}, {Kreidberg},
  {Kempton}, {McCullough}, {Fortney}, {Bean}, {Gillon}, {Homeier}, \&
  {Howard}}]{2014arXiv1403.4602K}
{Knutson}, H.~A., {Dragomir}, D., {Kreidberg}, L., {et~al.} 2014, ArXiv
  e-prints

\bibitem[{Leigh {et~al.}(2003)Leigh, {Collier Cameron}, Udry, Donati, Horne,
  James, \& Penny}]{2003MNRAS.346L..16L}
Leigh, C., {Collier Cameron}, A., Udry, S., {et~al.} 2003, \mnras, 346, L16

\bibitem[{{Lovis} \& {Fischer}(2010)}]{2011exop.book...27L}
{Lovis}, C. \& {Fischer}, D. 2010, {Radial Velocity Techniques for Exoplanets}
  (University of Arizona Press), 27--53

\bibitem[{{Luhman} \& {Jayawardhana}(2002)}]{2002ApJ...566.1132L}
{Luhman}, K.~L. \& {Jayawardhana}, R. 2002, \apj, 566, 1132

\bibitem[{{Madhusudhan} {et~al.}(2014){Madhusudhan}, {Crouzet}, {McCullough},
  {Deming}, \& {Hedges}}]{2014ApJ...791L...9M}
{Madhusudhan}, N., {Crouzet}, N., {McCullough}, P.~R., {Deming}, D., \&
  {Hedges}, C. 2014, \apjl, 791, L9

\bibitem[{Marley {et~al.}(1999)Marley, Gelino, Stephens, Lunine, \&
  Freedman}]{1999ApJ...513..879M}
Marley, M.~S., Gelino, C., Stephens, D., Lunine, J.~I., \& Freedman, R. 1999,
  \apj, 513, 879

\bibitem[{{Martins} {et~al.}(2013){Martins}, {Figueira}, {Santos}, \&
  {Lovis}}]{2013MNRAS.436.1215M}
{Martins}, J.~H.~C., {Figueira}, P., {Santos}, N.~C., \& {Lovis}, C. 2013,
  \mnras, 436, 1215

\bibitem[{Mayor {et~al.}(2003)Mayor, Pepe, Queloz, Bouchy, Rupprecht, {Lo
  Curto}, Avila, Benz, Bertaux, Bonfils, Dall, Dekker, Delabre, Eckert, Fleury,
  Gilliotte, Gojak, Guzman, Kohler, Lizon, Longinotti, Lovis, Megevand,
  Pasquini, Reyes, Sivan, Sosnowska, Soto, Udry, {van Kesteren}, Weber, \&
  Weilenmann}]{2003Msngr.114...20M}
Mayor, M., Pepe, F., Queloz, D., {et~al.} 2003, The Messenger, 114, 20

\bibitem[{Mayor \& Queloz(1995)}]{1995Natur.378..355M}
Mayor, M. \& Queloz, D. 1995, \nat, 378, 355

\bibitem[{{Miller-Ricci Kempton} \& Rauscher(2012)}]{2012ApJ...751..117M}
{Miller-Ricci Kempton}, E. \& Rauscher, E. 2012, \apj, 751, 117

\bibitem[{{Naef} {et~al.}(2004){Naef}, {Mayor}, {Beuzit}, {Perrier}, {Queloz},
  {Sivan}, \& {Udry}}]{2004A&A...414..351N}
{Naef}, D., {Mayor}, M., {Beuzit}, J.~L., {et~al.} 2004, A\&A, 414, 351

\bibitem[{{Oshagh} {et~al.}(2014){Oshagh}, {Santos}, {Ehrenreich},
  {Haghighipour}, {Figueira}, {Santerne}, \& {Montalto}}]{2014A&A...568A..99O}
{Oshagh}, M., {Santos}, N.~C., {Ehrenreich}, D., {et~al.} 2014, \aap, 568, A99

\bibitem[{Pepe {et~al.}(2002)Pepe, Mayor, Galland, Naef, Queloz, Santos, Udry,
  \& Burnet}]{2002A&A...388..632P}
Pepe, F., Mayor, M., Galland, F., {et~al.} 2002, A\&A, 388, 632

\bibitem[{{Pont} {et~al.}(2013){Pont}, {Sing}, {Gibson}, {Aigrain}, {Henry}, \&
  {Husnoo}}]{2013MNRAS.432.2917P}
{Pont}, F., {Sing}, D.~K., {Gibson}, N.~P., {et~al.} 2013, \mnras, 432, 2917

\bibitem[{{Redfield} {et~al.}(2008){Redfield}, {Endl}, {Cochran}, \&
  {Koesterke}}]{2008ApJ...673L..87R}
{Redfield}, S., {Endl}, M., {Cochran}, W.~D., \& {Koesterke}, L. 2008, \apjl,
  673, L87

\bibitem[{{Robin} {et~al.}(2003){Robin}, {Reyl{\'e}}, {Derri{\`e}re}, \&
  {Picaud}}]{2003A&A...409..523R}
{Robin}, A.~C., {Reyl{\'e}}, C., {Derri{\`e}re}, S., \& {Picaud}, S. 2003,
  \aap, 409, 523

\bibitem[{{Rodler} {et~al.}(2010){Rodler}, {K{\"u}rster}, \&
  {Henning}}]{2010A&A...514A..23R}
{Rodler}, F., {K{\"u}rster}, M., \& {Henning}, T. 2010, \aap, 514, A23

\bibitem[{Rodler {et~al.}(2013)Rodler, {K{\"u}rster}, {L{'o}pez-Morales}, \&
  Ribas}]{2013AN....334..188R}
Rodler, F., {K{\"u}rster}, M., {L{'o}pez-Morales}, M., \& Ribas, I. 2013,
  Astronomische Nachrichten, 334, 188

\bibitem[{Rodler {et~al.}(2012)Rodler, {Lopez-Morales}, \&
  Ribas}]{2012ApJ...753L..25R}
Rodler, F., {Lopez-Morales}, M., \& Ribas, I. 2012, \apjl, 753, L25

\bibitem[{Santos {et~al.}(2002)Santos, Mayor, Naef, Pepe, Queloz, Udry, Burnet,
  Clausen, Helt, Olsen, \& Pritchard}]{2002A&A...392..215S}
Santos, N.~C., Mayor, M., Naef, D., {et~al.} 2002, A\&A, 392, 215

\bibitem[{Santos {et~al.}(2013)Santos, Sousa, Mortier, Neves, Adibekyan,
  Tsantaki, {Delgado Mena}, Bonfils, Israelian, Mayor, \&
  Udry}]{2013A&A...556A.150S}
Santos, N.~C., Sousa, S.~G., Mortier, A., {et~al.} 2013, A\&A, 556, A150

\bibitem[{Schneider {et~al.}(2011)Schneider, Dedieu, {Le Sidaner}, Savalle, \&
  Zolotukhin}]{2011A&A...532A..79S}
Schneider, J., Dedieu, C., {Le Sidaner}, P., Savalle, R., \& Zolotukhin, I.
  2011, A\&A, 532, A79

\bibitem[{{Seager}(2010)}]{2010eapp.book.....S}
{Seager}, S. 2010, {Exoplanet Atmospheres: Physical Processes} (University of
  Arizona Press)

\bibitem[{{S{\'e}gransan} {et~al.}(2011){S{\'e}gransan}, {Mayor}, {Udry},
  {Lovis}, {Benz}, {Bouchy}, {Lo Curto}, {Mordasini}, {Moutou}, {Naef}, {Pepe},
  {Queloz}, \& {Santos}}]{2011A&A...535A..54S}
{S{\'e}gransan}, D., {Mayor}, M., {Udry}, S., {et~al.} 2011, \aap, 535, A54

\bibitem[{{Simpson} {et~al.}(2010){Simpson}, {Baliunas}, {Henry}, \&
  {Watson}}]{2010MNRAS.408.1666S}
{Simpson}, E.~K., {Baliunas}, S.~L., {Henry}, G.~W., \& {Watson}, C.~A. 2010,
  \mnras, 408, 1666

\bibitem[{{Snellen} {et~al.}(2014){Snellen}, {Brandl}, {de Kok}, {Brogi},
  {Birkby}, \& {Schwarz}}]{2014Natur.509...63S}
{Snellen}, I.~A.~G., {Brandl}, B.~R., {de Kok}, R.~J., {et~al.} 2014, \nat,
  509, 63

\bibitem[{Snellen {et~al.}(2010)Snellen, {de Kok}, {de Mooij}, \&
  Albrecht}]{2010Natur.465.1049S}
Snellen, I.~A.~G., {de Kok}, R.~J., {de Mooij}, E.~J.~W., \& Albrecht, S. 2010,
  \nat, 465, 1049

\bibitem[{{Snellen} {et~al.}(2010){Snellen}, {de Mooij}, \&
  {Burrows}}]{2010A&A...513A..76S}
{Snellen}, I.~A.~G., {de Mooij}, E.~J.~W., \& {Burrows}, A. 2010, \aap, 513,
  A76

\bibitem[{{Sobolev}(1975)}]{1975lpsa.book.....S}
{Sobolev}, V.~V. 1975, {Light scattering in planetary atmospheres} (Pergamon
  Press)

\bibitem[{{Stevenson} {et~al.}(2014){Stevenson}, {D{\'e}sert}, {Line}, {Bean},
  {Fortney}, {Showman}, {Kataria}, {Kreidberg}, {McCullough}, {Henry},
  {Charbonneau}, {Burrows}, {Seager}, {Madhusudhan}, {Williamson}, \&
  {Homeier}}]{2014arXiv1410.2241S}
{Stevenson}, K.~B., {D{\'e}sert}, J.-M., {Line}, M.~R., {et~al.} 2014, Science,
  346, 838

\bibitem[{Sudarsky {et~al.}(2003)Sudarsky, Burrows, \&
  Hubeny}]{2003ApJ...588.1121S}
Sudarsky, D., Burrows, A., \& Hubeny, I. 2003, \apj, 588, 1121

\bibitem[{{Traub} \& {Oppenheimer}(2010)}]{2011exop.book..111T}
{Traub}, W.~A. \& {Oppenheimer}, B.~R. 2010, {Direct Imaging of Exoplanets}
  (University of Arizona Press), 111--156

\bibitem[{{van Belle} \& {von Braun}(2009)}]{2009ApJ...694.1085V}
{van Belle}, G.~T. \& {von Braun}, K. 2009, \apj, 694, 1085

\bibitem[{{van Leeuwen}(2007)}]{2007A&A...474..653V}
{van Leeuwen}, F. 2007, \aap, 474, 653

\bibitem[{{Winn}(2010)}]{2011exop.book...55W}
{Winn}, J.~N. 2010, {Exoplanet Transits and Occultations} (University of
  Arizona Press), 55--77

\end{thebibliography}

\end{document}